\documentclass[aps,pra, preprint]{revtex4-1}
\usepackage{graphicx}
\usepackage{epsfig}
\usepackage{amsmath,bbm}
\usepackage{amssymb}
\usepackage{color}

\def\be{\begin{equation}}
\def\ee{\end{equation}}
\def\bea{\begin{eqnarray}}
\def\eea{\end{eqnarray}}
\def\bfk{{\bf k}}
\def\bfr{{\bf r}}

\begin{document}
\title{The Roles of Substrate vs Nonlocal Optical Nonlinearities in the Excitation of Surface Plasmons in Graphene}
\author{T. J. Constant}
\author{C. J. Tollerton}
\author{E. Hendry}
\affiliation{Electromagnetic Materials Group, Department of Physics, College of Engineering, Mathematics and Physical Sciences, University of Exeter, Exeter,
Devon EX4 4QL, UK}
\author{D. E. Chang}
\affiliation{ICFO Institut de Ci\`encies Fot\`oniques, Mediterranean Technology Park, 08860 Castelldefels (Barcelona), Barcelona, Spain.}
\date{\today}
\begin{abstract}
It has recently been demonstrated that difference frequency mixing (DFM) can generate surface plasmons in graphene\cite{Constant2015}. Here, we present detailed calculations comparing the contributions to this effect from substrate and from graphene nonlinearities. Our calculations show that the substrate (quartz) nonlinearity gives rise to a surface plasmon intensity that is around twelve orders of magnitude smaller than that arising from the intrinsic graphene response. This surprisingly efficient intrinsic process, given the centrosymmetric structure of graphene, arises almost entirely due to non-local contributions to the second order optical nonlinearity of graphene.

*Note: this is a preliminary manuscript written to coincide with the submission of author corresponding to ref. [\textit{Xiang and Gordon 2016 (Correspondence)}]. Due to time constraints, only the essential results are included. Further details will follow in subsequent version(s).
\end{abstract}
\maketitle
There has been substantial recent interest in the excitation and manipulation of surface plasmons in graphene \cite{Fei2012,Chen2012b,Wagner2014,Jadidi2016,Luxmoore2014,Yan2013}. These surface electromagnetic waves are very strongly bound to the graphene, confining the electromagnetic energy in a near surface region that is around two orders of magnitude thinner than the wavelength of light for the equivalent frequency \cite{Abajo2014,koppensnl}. While the effects of these surface modes can be indirectly seen in nanostructured graphene (e.g. in graphene nanoribbons \cite{Yan2013,Luxmoore2014}), there has been a push to find ways to excite and manipulate them, first via near-field scattering techniques\cite{Chen2012,Fei2012}, and more recently using resonant metal antennas \cite{Alonso-Gonzalez2014}.

Most recently, it has been shown that one can use optical nonlinearities to excite wavevector defined surface plasmons in graphene \cite{Constant2015,Jadidi2016,Yao2014}. In ref. \cite{Constant2015}, we presented results showing that surface plasmon modes could be excited in graphene via nonlinear wave mixing. In the reported excitation scheme, nonlinear mixing of two visible-frequency light sources is used to generate an evanescent difference frequency field with sufficient momentum to couple to the surface plasmons in graphene. This effect has been attributed to the intrinsic nonlinear optical response of graphene itself\cite{Constant2015}. This might be viewed as rather unexpected, since graphene is centrosymmetric and, within the dipole approximation, should have no second order nonlinearity \cite{Boyd2008}. We attributed the effect to a spatially nonlocal interactions in graphene, which has been predicted theoretically to enable second-order response \cite{Yao2014}. However, it has been pointed out [\textit{Xiang and Gordon, 2016 (Correspondence)}] that another excitation mechanism is possible, via the nonlinear response of the dielectric substrate (in ref. \cite{Constant2015}, quartz). The idea here is a nonlinear polarization of the underlying substrate couples, through the linear density of states, to surface plasmons in the graphene. In this process, the optical nonlinearity of the graphene itself plays no role in the mechanism, and only the high linear density of states near the surface plasmon resonance supports resonant field enhancement. 

It is clearly important to establish the efficiency of the two mechanisms described above. In this letter, we present detailed calculations of the two processes above. First, we present a self-consistent derivation which describes the linear response of a graphene layer with a nonlinear substrate, including both realistic plasmon losses and phase matching. Using this model with the experimental parameters from \cite{Constant2015} and taking an approximate value of for the nonlinear substrate susceptibility of $0.6 \mathrm{pm V^{-1}}$ (Quartz) \cite{Shoji2002}, we find a completely different plasmon excitation spectrum than the case where graphene itself provides the nonlinearity. In particular, because the pump and probe beam properties are chosen to phase-match to graphene plasmons, their difference frequency-wavevector relation is far from phase-matched with propagating fields in quartz. This highly suppresses the nonlinear field generated, and consequently the induced plasmon field. We predict an electric field amplitude in the graphene layer of $\approx 15 \mathrm{V m^{-1}}$. We then include the optical nonlinearity of the graphene itself, derived from first principles following Yao \textit{et.~al}\cite{Yao2014}, and find and electric field in the graphene layer that is more than six orders of magnitude greater, suggesting the generation of surface plasmons occurs almost entirely due to non-local contributions to the second order optical nonlinearity of the graphene itself.

\begin{figure}
\includegraphics[width=0.5\linewidth]{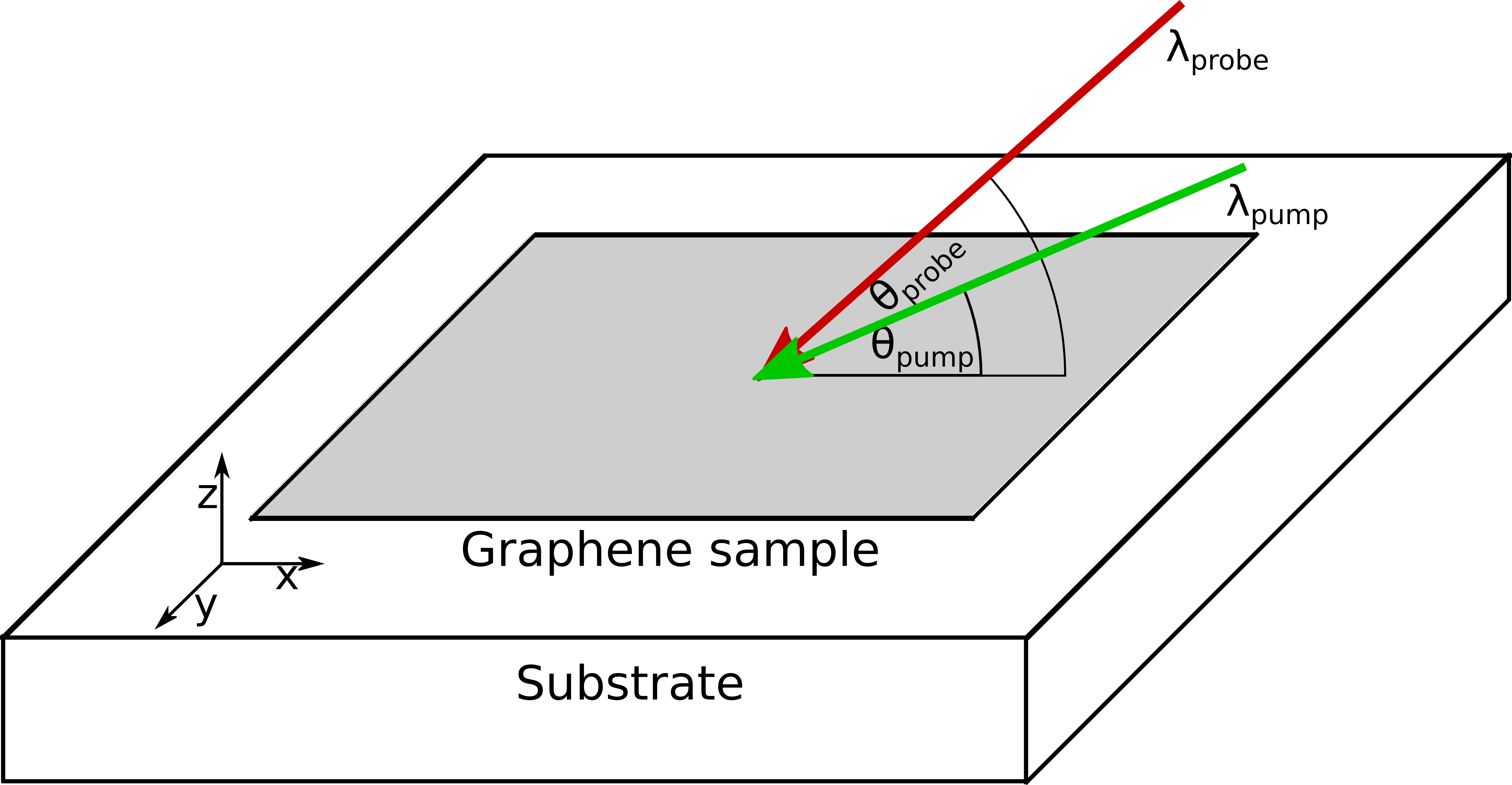}
\caption{The coordinate system under consideration. Two beams of wavelengths $\lambda_{pump}$ and $\lambda_{probe}$ are incident on a graphene sample on a quartz substrate, impinging with angles $\theta_{pump}$ and $\theta_{probe}$, respectively.}
\end{figure}

\section{Substrate Contribution}

We begin by considering separately the nonlinear response of the underlying substrate (i.e. for now, we consider the graphene to be a purely linear material). The geometry of the sample and incident fields is defined in figure 1. The analysis significantly simplifies if the nonlinear polarization is generated far from a phase matching condition of the bulk, and depletion can be ignored, as should be the situation in experiment. In this case, the pump and probe fields generate a polarization in quartz of
\be P_3(\bfr,t)=\frac{1}{2}\epsilon_0\chi^{(2)}e^{i(\bfk_{T1}-\bfk_{T2})\cdot\bfr-i(\omega_1-\omega_2)t}E_{T1}E_{T2}^{\ast}, \ee
where $\bfk_{Ti}$,$E_{Ti}$ denotes the wavevector and field amplitudes of the pump~($i=1$) and probe~($i=2$) fields on the transmitted~(substrate) side. With the assumptions above, $\bfk_{Ti}$,$E_{Ti}$ can be calculated purely from the linear optical Fresnel coefficients and Snell's Law. The subscript $i=3$ indicates quantities corresponding to the difference frequency signal at $\omega_3=\omega_1-\omega_2$. As charge density waves in graphene are driven by an electric field, we must relate the nonlinear polarization to the field generated in the quartz, which satisfies the wave equation
\be -\nabla^2 E_{3s}+\frac{\epsilon_3}{c^2}\frac{\partial^2 E_{3s}}{\partial t^2}=-\mu_0 \frac{\partial^2 P_3}{\partial t^2}. \ee
Here, the subscript ``s'' denotes that this is an effective source field that will later drive a response in graphene~(and thus the subscript provides a distinction from the resulting plasmon field). Also, $\epsilon_3=\epsilon(\omega_3)$ indicates the permittivity of quartz evaluated at the difference frequency. Due to the plane-wave nature of $P_3$, $E_{3s}$ takes on the same spatial and frequency dependence. In our regime of interest, the spatial derivative of the field, $|\nabla^2 E_{3s}|=|\bfk_{T1}-\bfk_{T2}|^2 E_{3s}^2$, is significantly larger than the time derivative. This is because the pump and probe fields are chosen to phase-match with surface plasmons in graphene (i.e. the wavevectors are much larger than free-space fields of the same frequency). Thus the field amplitude created by the nonlinear polarization is well-approximated by
\be E_{3s}\approx \frac{(\omega_1-\omega_2)^2}{2c^2 |\bfk_{T1}-\bfk_{T2}|^2}\chi^{(2)}|E_{T1}E_{T2}^{\ast}|. \ee
In particular, it should be noted that a large wavevector mismatch results in a strong suppression of the field. Thus far we have neglected the vector nature of the field $E_{3s}$, which depends on the tensor nature of the nonlinear susceptibility $\chi^{(2)}$ and the polarizations of the pump and probe fields. To simplify the discussion, we will assume the scenario which produces the highest field, i.e. in which $E_{3s}$ is completely polarized along $\hat{x}$~(parallel to the graphene sheet) so that it maximally drives a charge density wave in graphene. As we see below, even in this best case scenario, the generated field is rather small.

We now calculate the resulting plasmon field amplitude. Since the hypothesis is that the nonlinear response is completely within the substrate and its role is to provide an effective source field $E_{3s}$, the remaining part of the calculation is completely linear in its nature. Using the same conventions as Fig. S6 in the Supplementary Information of ref.~\cite{Constant2015}, we take ``reflected'' and ``transmitted'' field components of unknown amplitude, which correspond to the plasmon fields on the vacuum and substrate sides. The wavevector along $\hat{x}$ for these fields is equal to $k_{3x}=k_{T1x}-k_{T2x}$, while the perpendicular components must satisfy the respective dispersion relations for each side, \textit{e.g.}, $k_{T3z}^2=\epsilon_3 (\omega_3/c)^2-k_{3x}^2$. The two unknown field amplitudes can be readily solved by enforcing electromagnetic boundary conditions at the vacuum-graphene-quartz interface, which yields the following parallel-field component on the substrate side, evaluated at the graphene layer~$(z=0)$,
\be E_{pl}=-E_{3s}\frac{(c\epsilon_0+\sigma_3 \sin\theta_3)\sin\phi_3}{c\epsilon_0\sin\phi_3+\sin\theta_3(c\epsilon_0\sqrt{\epsilon_3}+\sigma_3\sin\phi_3)}. \ee
Here, $\sigma_3=\sigma^{(1)}(\omega_3)$ is the linear conductivity of graphene evaluated at frequency $\omega_3$. A couple of limits are worth mentioning. First, if there are no losses in quartz or graphene, using Eq. (S8) in ref.~\cite{Constant2015} for the linear conductivity of graphene and setting the denominator to zero in the above equation recovers the linear dispersion relation of graphene plasmons (Eq. (S9) in ref.~\cite{Constant2015}). Furthermore, setting $k_{3x}$ to satisfy this dispersion relation in the presence of losses in graphene~(but with no losses in quartz), one finds $E_{pl}\approx -iE_{3s}\epsilon_3 Q/(1+\epsilon_3)$, where $Q=\omega_3/\gamma$ is the plasmon quality factor. This equation thus recovers the physically reasonable expectations that the field is enhanced when the source phase matches to the dispersion relation for graphene plasmons, and in this case, the generated plasmon field grows with decreasing losses.

More generally, we can numerically evaluate $E_{pl}$ for the conditions specific to the experiment\cite{Constant2015}, where the permittivity $\epsilon_3$ of quartz is modified and becomes complex due to coupling with phonons~(see SI, ref.~\cite{Constant2015}). As a specific example, we take the parameters from ref.~\cite{Constant2015}, where $\theta_{\textrm{\footnotesize{pump}}}=50^{\circ},\theta_{\textrm{\footnotesize{probe}}}=70^{\circ}$, and where a peak differential reflection signal was observed at $\lambda_{\textrm{\footnotesize{probe}}}=615$~nm, $\lambda_{\textrm{\footnotesize{pump}}}\approx 571$~nm. Using incident intensities  $I_{\textrm{\footnotesize{pump}}}\approx 10^{13}\,\mathrm{~W m^{-2}}$ and $I_{\textrm{\footnotesize{probe}}}\approx 10^{11}\,\mathrm{~W m^{-2}}$ (i.e. comparable to experiment\cite{Constant2015}, and taking a value of $\chi^{(2)}=0.6~\mathrm{pm V^{-1}}$ for quartz \cite{Shoji2002}, we find that $E_{pl}\approx 15\,\mathrm{~V m^{-1}}$.

\section{Nonlocal contribution from graphene}
Under certain circumstances outwith the dipole-approximation, second order nonlineararities in centrosymmetric materials are predicted, and can even provide rather large second order nonlinear effects \cite{Mikhailov2011,Lobet2015ProbingSieve}. In graphene, such nonlocal contributions can be surprisingly large \cite{Mikhailov2011,Manzoni2015a,Lobet2015ProbingSieve}, arising from an inverse Fermi momentum that is significantly larger than in typical conductors\cite{Manzoni2015a}. In this section, we include the nonlocal optical nonlinearity of the graphene itself, derived from first principles following Yao \textit{et.~al}\cite{Yao2014}. Note that, for consistency, our derivation is carried out in SI units, as opposed to the ESU units originally used in \cite{Yao2014}.

\begin{figure}
\includegraphics[width=0.5\linewidth]{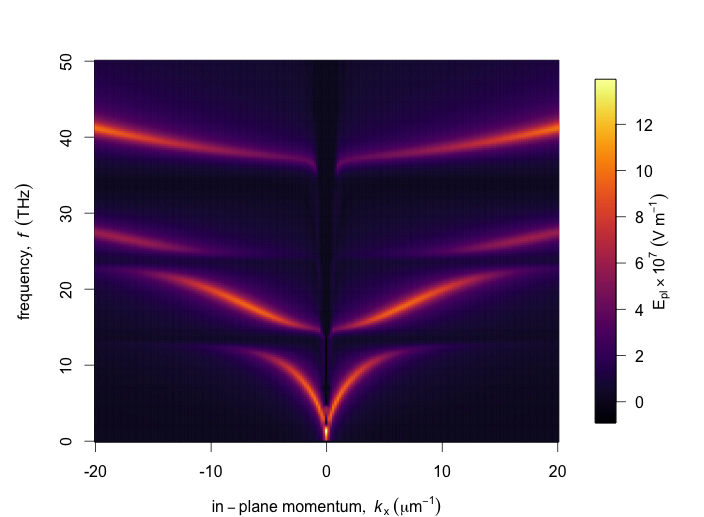}
\caption{Plasmon electric field strength as a function of in-plane wavevector and frequency for the case of a nonlinear response from the graphene. The simple plasmon dispersion is modified in the presence of substrate phonons, causing four branches for hybrid surface plasmon-phonon excitations.}\label{fig:yao-plasmon-dispersion}
\end{figure}

\begin{figure}
\includegraphics[width=0.5\linewidth]{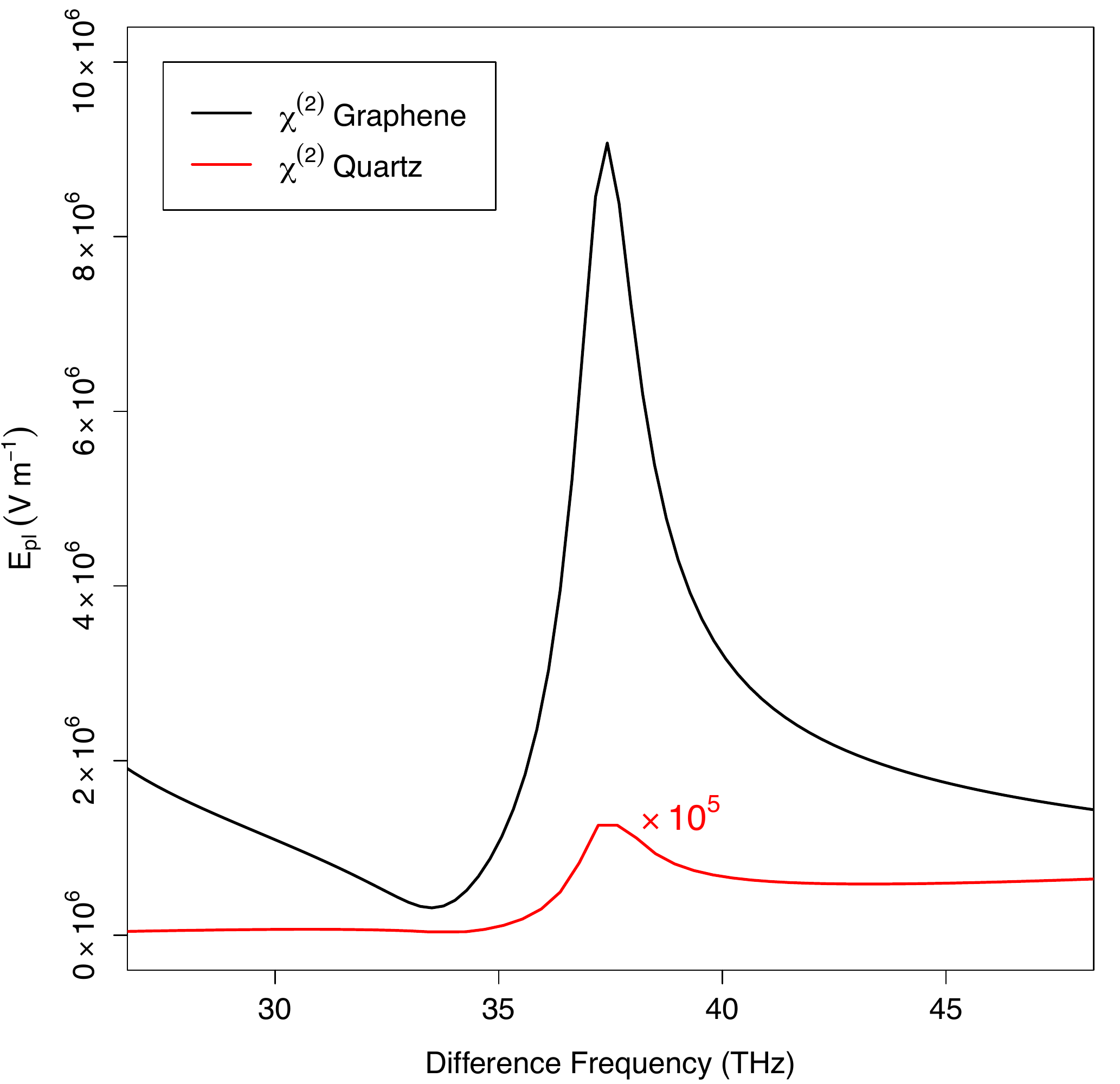}
\caption{Comparison of the electric field strength in the surface plasmon as a function of difference frequency, for the cases of the nonlinear response of graphene (black line) and via the nonlinear response of the substrate (red line). The electric field strength for the nonlinear substrate response (peak $E_{pl}\approx 15\,\mathrm{V m^{-1}}$), has been multiplied by $\times 10^5$ to be visible on this scale. The two mixing fields are incident with angles $\theta_{\textrm{\footnotesize{pump}}}=50^{\circ},\theta_{\textrm{\footnotesize{probe}}}=70^{\circ}$.}\label{fig:comparsion-field-plot}
\end{figure}

Firstly we take the following plasmon dispersion,
\begin{equation}
D(\omega,q)= 4\pi\tilde{\chi}  + \frac{\epsilon_1}{p_1} + \frac{\epsilon_2}{p_2},
\end{equation}
with the condition that on resonance $\mathrm{Real}(D) = 0$. Here $\tilde{\chi}$ 
is the linear susceptibility of the graphene, $\epsilon_{1,2}$ are the frequency dependent permittivities of the dielectrics either side of the graphene and $p_{1,2}$ are the penetration wavevectors given by $p_{1,2} = \sqrt[2]{q^2-\epsilon_{1,2}\frac{\omega^2}{c^2}}$. We take $\epsilon_{1}=1$, while $\epsilon_{2}(\omega)$ is the frequency dependent permittivity of quartz, taken from ref.~\cite{Luxmoore2014}. Both the first and second order susceptibility of graphene are calculated from the iterative solutions of the density matrix and interaction Hamiltonian;
\begin{equation}
\rho_{nm}^{(N)} = \int^t-\frac{i}{\hbar}[\hat{H}_{int}(t'),\rho^{(N-1)}]_{nm}exp[(i\omega_{nm}+\gamma_{nm})(t'-t)]dt'.
\end{equation}
Here $\rho^{(N)}_{nm}$ is the density matrix element corresponding to two arbitrary electron states $n$ and $m$, $\omega_{nm}$ is the energy difference between the two states and $\gamma_{nm}$ is the scattering rate between them, we henceforth model this as a uniform scattering rate of $\gamma$. The interaction Hamiltonian is modeled as
\begin{equation}
\hat{H}_{int} = iev_f\hat{\vec{\sigma}}.\vec{E},
\end{equation}
with
\begin{equation}
\hat{\vec{\sigma}} = \hat{\sigma}_a\hat{x}+\hat{\sigma}_y\hat{y},
\end{equation}
where $\hat{\sigma}_x$ and $\hat{\sigma}_y$ are the Pauli spin matrices.The second order susceptibility for a DFM process of two incoming beams is found to be
\begin{equation}
\begin{aligned}
\chi^{(2)}_{ijk}=&\\
&\left[\frac{c^2}{10^7}\right]\frac{g e^2}{\hbar^2\omega_a\omega_b}\int\int\frac{d^2k_1}{(2\pi)^2}\left(\left(\frac{f(k_1)-f(k_3)}{\omega_{31} -\omega_b -i\gamma}+\frac{f(k_1)-f(k_2)}{-\omega_{21}+\omega_a-i\gamma}\right)\frac{\mu^i_{32}v^j_{31}v^k_{12}}{\omega_{32}-\omega-i\gamma}-\right.\\
&\hspace{4em}\left.\left( \frac{f(k_1)-f(k_3)}{\omega_{31}-\omega_b-i\gamma}+\frac{f(k_{2'})-f(k_3)}{-\omega_{32'} +\omega_a-i\gamma}\right)\frac{\mu^i_{2'1}v^j_{31}v^k_{2'3}}{\omega_{2'1}-\omega-i\gamma}\right).\label{eq:yao-integral}
\end{aligned}
\end{equation}
Here $ \chi^{(2)}_{ijk}$ denotes the surface susceptibility of the graphene. One could in principal use this full expression to find the nonlocal nonlinearity of graphene. However, we instead use the same limits of Yao \textit{et.~al} \textit{et.~al}\cite{Yao2014} (i.e.  $\omega >> v_fq, \gamma$, $k_bT\Rightarrow0$ and $\omega_a \sim \omega_b$). Under these assumptions, equation \ref{eq:yao-integral} can be approximated to
\begin{equation}
\chi^{(2)}_{xxx} \sim \frac{e^3}{8\pi\hbar^2}\frac{g}{q\omega_1\omega}\left(\frac{\pi}{2}+\arctan\left(\frac{\omega_1-2v_fk_f}{\gamma}\right)\right)\left[\frac{c^2}{10^7}\right].\label{eq:yao-approximated-chi2}
\end{equation}
Note that the $1/q$ dependence in equation \ref{eq:yao-approximated-chi2}, which arises under the approximations applied above, is unphysical for $q$ tending to zero. This gives an erroneous divergence in the calculation for small $q$. Nevertheless, we believe the approximations used by Yao \textit{et.~al} \cite{Yao2014} are valid in the vicinity of the plasmon resonance. Note that the susceptibility itself does not contain any resonance for the surface plasmon; this resonance is observed in the field generated from this polarization, given by
\begin{equation}
E_{pl}=-\frac{4\pi}{D(\omega,q)}\chi^{(2)}_{xxx}E_{T1}E_{T2}^*.
\end{equation}
This equation allows us to obtain an order of magnitude estimate for the generated electric field of the surface plasmon. Using the same parameters from section I above, we calculate the electric field in the graphene layer. This is plotted in figure \ref{fig:yao-plasmon-dispersion} as a function of difference frequency and in-plane wavevector. We see large enhancements to the electric field amplitude near the plasmon resonance conditions. Note the horizontal asymptotes correspond to the surface phonon frequencies of quartz \cite{Luxmoore2014}. In figure \ref{fig:comparsion-field-plot} we compare the electric field generated as a function of difference frequency, for both the models from sections I and II ($\theta_{\textrm{\footnotesize{pump}}}=50^{\circ},\theta_{\textrm{\footnotesize{probe}}}=70^{\circ}$). Near the plasmon resonance condition, we find an electric field strength ~$9 \times 10^6 \mathrm{V m^{-1}}$ for the model which includes the graphene nonlinearity, i.e. around six orders of magnitude grater than found without the graphene nonlinearity in section I. 

\section{Conclusions}
We present detailed calculations comparing the contributions to surface plasmon generation via second order substrate and graphene nonlinearities. We find the efficiency generation via a quartz substrate nonlinearity is negligibly small, while the graphene nonlinearity gives rise intensity that is around twelve orders of magnitude larger. This surprisingly efficient process, given the centrosymmetric structure of graphene, arises almost entirely due to non-local contributions to the second order optical nonlinearity of graphene. We suggest that systems where the substrate composition and orientation are optimized could give larger contributions from the substrate, and may be an interesting route toward increasing the efficiency of the excitation process.

\bibliography{Mendeley}

\end{document}